\documentclass[pra,twocolumn,superscriptaddress]{revtex4-1}
\usepackage{mathrsfs}
\usepackage{amsmath}
\usepackage{amssymb}
\usepackage{graphicx}
\usepackage{esint}
\usepackage{times}

\begin{document}
\title{Decoherence-free propagation and ramification of a solitary pulse
in a superconducting circuit}
\author{Yibo Gao}
\affiliation{Beijing University of Technology, Beijing, China}
\author{Shijie Jin}
\affiliation{Beijing University of Technology, Beijing, China}
\author{Hou Ian}
\affiliation{Institute of Applied Physics and Materials Engineering, University
of Macau, Macau, China}
\affiliation{UMacau Research Institute, Zhuhai, Guangdong, China}
\email{houian@um.edu.mo}

\begin{abstract}
Using a microscopic master equation to account for the environmental
effects, we compute the decoherence culminated during the propagation
of a microwave pulse of arbitrary shape through a superconducting
qubit. It is shown that the qubit decoherence vanishes and the pulse
shape remains absorption-free when the latter adopts a soliton shape
with $n\pi$ area. Otherwise, the environmental feedback decelerates
the velocity of the soliton envelop and induces an monotonic increase
of phase in the microwave. A pulse of non-$n\pi$ area thus ramifies
into a transparent part that travels decoherence-free at incident
velocity and a slowing part that decays through space. The ramification
explains the environmental origin of pulse splitting observed in self-induced
transparency.
\end{abstract}
\maketitle

\section{Introduction}

Superconducting circuit systems comprising one or a few qubits are
controlled by microwave pulses. When coupled with a cavity field residing
in a coplanar stripline resonator, these qubits form a pulse-controlled
circuit quantum electrodynamic (cQED) system~\citep{wallraff04,blais04},
which serves as the foundation of solid-state entanglement generation~\citep{neeley10,eichler12}
and quantum computing~\citep{marian11,lucero12}. For examples, resonant
and dispersive square pulses with appropriate lengths are transmitted
to read out qubit states~\citep{mallet09}; Gaussian pulses are used
to perform $x$- and $y$-rotations to achieve GHZ states~\citep{dicarlo10}.
In general, the enveloping shapes of the microwave pulses are proven
to be pivotal to the desired operations on the qubits~\citep{motzoi09,chow10}.

A natural question to ask then is whether microwave pulses can be
used to eliminate decoherence. There had been multiple approaches
that targets qubit decoherence. First, the problem is algebraically
approached through decoherence-free subspaces~\citep{lidar98}, which
is implemented in a superconducting circuit using Purcell effect~\citep{gambetta11}.
The second approach is device-based, where the anharmonicity of the
qubit is fine tuned through, for instance, the introduction of transmon~\citep{koch07}.
The latter can prolong the decoherence times to over 2$\mu$s~\citep{schreier08}.
In this article, we take a third quantum-optical approach to determine
the circumstance when the decoherence of a qubit can be removed through
the interaction of a solitary microwave pulse with the qubit in a
superconducting circuit.

Since a superconducting qubit when biased to an optimal operation
point (e.g. reduced flux $f=1/2$) is essentially a two-level atom,
many optical effects induced by atom-field interactions, such as electromagnetically
induced transparency~\citep{ian10} and parametric amplificataion~\citep{wen18},
also apply to qubits on a superconducting circuit~\citep{jqyou11}.
The closest analogous study to the scenario considered here that occurs
in natural atomic systems is the propagation of narrow pulses in resonant
atomic media~\citep{lamb71}, for which the effect of self-induced
transparency (SIT) is a prime manifestation~\citep{mccall67,mccall69}.
Nevertheless, the relative scale of a microwave pulse with respect
to a superconducting qubit is incommensurable to that of an optical
pulse with respect to an atomic medium. For example, SIT regards an
optical narrow pulse as those widths at half height being much shorter
than the length of the atomic medium (e.g. 7 nsec pulses in a 1 mm
Rb-sample at density $n=10^{11}\mathrm{cm}^{-3}$~\citep{slusher72}).
In contrast, a microwave pulse at nanosecond range~\citep{mallet09}
translates to a length of centimeter range when propagating in a silicon-substrated
circuit and facing a qubit typically measures at only 300$\mu$m~\citep{wen18}.
The scenario is illustrated in Fig.~\ref{fig:model}.

Therefore, the scenarios of SIT and the microwave propagation through
a qubit are analogous although the roles of the field and the medium
are switched due to their relative length scales. At resonance, the
traveling pulse is similarly described by an inhomogeneous Maxwell
equation, with the inhomogeneous term contributed by the polarization
of the medium, i.e. the density matrix of the qubit. Nevertheless,
we introduce a microscopic adiabatic master equation approach to depict
the time evolution of the qubit, taking all decoherence effects into
account, in contrast to past analyses where $T_{1}$ and $T_{2}$
relaxation times are assumed infinite to simplify calculations~\citep{mccall69,lamb73}.
We analytically solve the master equation to trace the qubit decay
through a complex decoherence factor, whose real and imaginary parts
correspond to the dephasing and the longitudinal relaxation susceptible
by the qubit. A hypersecant solution is found to associate with a
microwave propagation of no absorption by the qubit. More importantly,
during the transparent propagation, the culminated longitudinal relaxation
of the qubit also vanishes while the transverse relaxation is determined
by an integral formula. Zero-dephasing evolution during propagation
is theoretically obtainable when the spectral distribution of the
reservoir becomes orthogonal to the sinusoids of the pulse phase.
The absorption-free propagation is similar to SIT but the decoherence-free
propagation has not been discovered before.

Consequently, solving for the envelop of the pulse shows that a non-transparent
pulse experiences a reduced propagation velocity while its energy
is absorbed. Hence, an arbitrary solitary pulse is ramified into a
hypersecant part which travels at the incident velocity and a remainder
part whose travel is dragged by environmental feedbacks. Therefore,
the microscopic approach here points out the thermal environmental
origin of pulse splitting observed in SIT~\citep{mccall69,slusher72}
and provides clues for decoherence control using microwave pulses
in superconducting circuits. We begin the study by describing the
qubit-pulse interaction model in Sec.~\ref{sec:model} and follow
with the derivation of the decoherence factor in Sec.~\ref{sec:decoh_factor}.
The full solutions of the pulse envelop and phase are presented in
Sec.~\ref{sec:prop_and_ram} along with the discussion of pulse ramifications,
before a conclusion is given in Sec.~\ref{sec:conclusion}.

\section{Qubit-pulse interaction in thermal environment\label{sec:model}}

\begin{figure}
\includegraphics[clip,width=8.5cm]{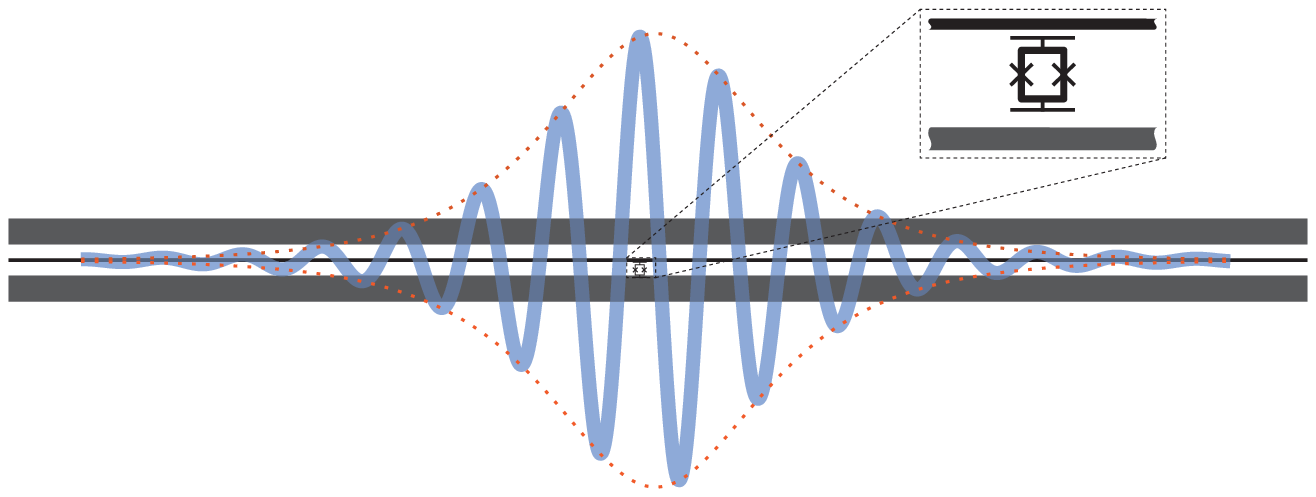}

\caption{Illustration of the model (not drawn to scale): a superconducting
qubit (detail circuit model shown in the inset, typical dimension
measured at $300$$\mu$m for a transmon qubit~\citep{wen18}) in
the vicinity of a coplanar waveguide (black core line surrounded by
dark gray ground strips) facing a traveling microwave field (envelop
drawn in orange, wavelength measured at 2~cm for a typical 5~GHz
microwave on a silicon substrate).~\label{fig:model}}
\end{figure}

We begin the derivation by assuming the electric field of an incident
microwave pulse take the form 
\begin{equation}
E(x,t)=\mathcal{E}(x,t)\cos\left[\varphi(x,t)-kx+\omega t\right]\label{eq:elec_field}
\end{equation}
where $\mathcal{E}(x,t)$ and $\varphi(x,t)$ denote, respectively,
its envelop and phase during its traveling along a waveguide. $\omega$
denotes the frequency and $k$ the associated wavevector of the carrier
wave. In Eq.~(\ref{eq:elec_field}), $x$ and $t$ denote laboratory
spatial and time coordinates but they are customarily compressed~\citep{lamb71,basov66}
into the single variable $\tau=t-x/v$ of local time under the reference
frame that travels along with the wavefront. Since dispersive effects
are not considered, the velocity $v$ of the wavefront is assumed
constant $v=\omega/k$ throughout the propagation and the electric
field becomes a single-variant function $E(\tau)=\mathcal{E}(\tau)\cos\left[\varphi(\tau)+\omega\tau\right]$
undering the traveling reference frame. The system, illustrated in
Fig.~\ref{fig:model}, is described by the time-dependent Hamiltonian
$(\hbar=1)$
\begin{equation}
H_{\mathrm{S}}(\tau)=\frac{\omega_{0}}{2}\sigma_{z}+\mu E(\tau)\left[\sigma_{+}+\sigma_{-}\right]
\end{equation}
where $\omega_{0}$ and $\mu$ are the transition frequency and the
effective dipole moment, respectively, of the qubit. In other words,
$E(\tau)$ is non-zero during the propagation of the pulse through
the qubit; otherwise, $E(\tau)$ vanishes, letting the qubit evolve
freely and the pulse travel freely.

This semiclassical Hamiltonian is diagonalizable through the dressed
states
\begin{align}
\left|\nu_{+}(\tau)\right\rangle  & =e^{-i\varphi(\tau)}\cos\theta(\tau)\left|e\right\rangle -\sin\theta(\tau)\left|g\right\rangle ,\label{eq:nu_+}\\
\left|\nu_{-}(\tau)\right\rangle  & =e^{-i\varphi(\tau)}\sin\theta(\tau)\left|e\right\rangle +\cos\theta(\tau)\left|g\right\rangle \label{eq:nu_-}
\end{align}
for the eigenvalues $\Omega_{\pm}=\pm\sqrt{\delta^{2}+(\mu\mathcal{E})^{2}}$
in the rotating frame $e^{i\omega\tau}$ of the microwave carrier.
The transformation angle 
\begin{equation}
\theta(\tau)=\frac{1}{2}\tan^{-1}\frac{\mu\mathcal{E}}{\delta}
\end{equation}
 depends on the qubit-field detuning $\delta=\omega_{0}-\omega$.
When the pulse is not overlapping with the qubit, the two dressed
states assume the asymptotic $\left|e\right\rangle $ or $\left|g\right\rangle $
of the bare states.

The environmental effects to the qubit are modeled on a multi-mode-resonator
bath with free Hamiltonian $H_{\mathrm{B}}=\sum_{j}\omega_{j}a_{j}^{\dagger}a_{j}$.
Paired with the diagonalized $H_{\mathrm{S}}^{\prime}=\Omega_{+}\left|\nu_{+}\right\rangle \left\langle \nu_{+}\right|-\Omega_{-}\left|\nu_{-}\right\rangle \left\langle \nu_{-}\right|$,
the universe has the Hamiltonian $H=H_{\mathrm{S}}^{\prime}+H_{\mathrm{B}}+H_{\mathrm{I}}$,
where the system-bath coupling is the tensor product 
\begin{equation}
H_{\mathrm{I}}=h_{\mathrm{S}}\otimes h_{\mathrm{B}}=\left[\left|e\right\rangle \left\langle g\right|+\left|g\right\rangle \left\langle e\right|\right]\otimes\sum_{j}g_{j}(a_{j}+a_{j}^{\dagger}).\label{eq:int_H}
\end{equation}
Since the system eigenstates does not remain static but rather follow
the change in the amplitude of the microwave pulse, the basis for
system-bath coupling should align with the time-dependent basis in
Eqs.~(\ref{eq:nu_+})-(\ref{eq:nu_-}) to take into account of dressed
relaxations~\citep{ian10,wilson07}. While the dressed system and
the bath mutually affect each other during the process of qubit-field
interaction, the evolution is adiabatic on the system side~\citep{BREUER}.
That is, following the reference frame of the transient pulse, we
consider the evolution of the system density matrix $\rho^{\prime}=U_{\mathrm{ad}}\rho U_{\mathrm{ad}}^{\dagger}$
under the interaction picture, where
\begin{multline}
U_{\mathrm{ad}}(\tau)=\left|\nu_{+}(\tau)\right\rangle \left\langle \nu_{+}(0)\right|e^{-i\phi_{+}(\tau)}\\
+\left|\nu_{-}(\tau)\right\rangle \left\langle \nu_{-}(0)\right|e^{-i\phi_{-}(\tau)}\label{eq:U_ad}
\end{multline}
denotes the unitary matrix during the adiabatic evolution up to time
$\tau$. In Eq.~(\ref{eq:U_ad}), 
\begin{equation}
\phi_{\pm}(\tau)=\int_{\tau_{0}}^{\tau}ds\left[\Omega_{\pm}(s)-i\left\langle \nu_{\pm}(s)|\dot{\nu}_{\pm}(s)\right\rangle \right]
\end{equation}
expresses the total phase, i.e. both the dynamic (the first term)
and the geometric phase (the second term), culminated in each dressed
state.

Our considerations begin with the Liouville equation for the system
density matrix $\rho$ in the Schroedinger picture, which reads 
\begin{equation}
\frac{d\rho^{\prime}}{d\tau}=-\int_{\tau_{0}}^{\tau}ds\left\langle \left[H_{I}(\tau),\left[H_{I}(s),\rho^{\prime}(\tau)\otimes\tilde{\rho}\right]\right]\right\rangle \label{eq:Liouville_eqn}
\end{equation}
where the integral corresponds to the first nontrivial term in the
perturbative expansion of time-ordered evolution of the universe.
The density matrix $\tilde{\rho}$ of the bath will be partial-traced
out when taking the ensemble average. This integral represents the
memory effect of the bath onto the system during the propagating of
the pulse through the qubit under the Born-Markov approximation. Expanding
the double commutator will give four terms involving both $\tau$
and $\tau'$. Those related to the bath part only involves double-time
correlations when taking the trace and read~\citep{BREUER}
\begin{multline}
\left\langle h_{\mathrm{B}}(\tau)h_{\mathrm{B}}(\tau-\tau')\right\rangle =\\
\left\langle h_{\mathrm{B}}(\tau)h_{\mathrm{B}}(\tau-\tau')\right\rangle ^{\ast}=\sum_{j}g_{j}^{2}e^{-i\omega_{j}\tau'}.\label{eq:correl}
\end{multline}
Those related to the system can be considered separately. Following
the method~\citep{albash12}, the memory effect can be recorded by
reversing the direction of time ($\tau'\to\tau-\tau'$) such that
$U_{\mathrm{ad}}(\tau-\tau')=\exp\left\{ i\tau'H_{S}(\tau)\right\} U_{\mathrm{ad}}(\tau)$.

\section{Decoherence factor\label{sec:decoh_factor}}

Expanding the commutators and tracing out the bath operators in the
Liouville equation yields the microscopic master equation in the Lindblad
form
\begin{multline}
\frac{d\rho}{d\tau}=-i\left[H_{\mathrm{S}},\rho\right]+\gamma(\Omega)\sin^{2}\varphi\\
\times\left[\hat{\sigma}_{-}\rho\hat{\sigma}_{+}-\frac{1}{2}\left\{ \hat{\sigma}_{+}\hat{\sigma}_{-},\rho\right\} \right]\label{eq:master_eqn}
\end{multline}
after converting to the Schroedinger picture. The Pauli matrices are
hatted to indicate that the dressed basis is assumed, e.g. $\hat{\sigma}_{x}=\left|\nu_{+}(\tau)\right\rangle \left\langle \nu_{-}(\tau)\right|+\left|\nu_{-}(\tau)\right\rangle \left\langle \nu_{+}(\tau)\right|$.
\begin{equation}
\gamma(\Omega)=2\pi\sum_{j}g_{j}^{2}\delta(\omega_{j}-\Omega)\label{eq:gamma}
\end{equation}
denotes the spectral density distribution of the bath stemming from
the integration, which is essentially the Fourier transform of Eq.~(\ref{eq:correl}).
The detailed derivations of Eq.~(\ref{eq:master_eqn}) is given in
Appendix A The effect of an incident pulse on the qubit is reflected
in its polarization $P(\tau)=\mu\mathrm{tr}\left\{ (\hat{\sigma}_{x})\rho(\tau)\right\} $
as a time-dependent response to the pulse, where the trace is taken
over the dressed system basis. From Eq.~(\ref{eq:master_eqn}), one
can derive that $P(\tau)=\mu\mathcal{F}\exp i\{\varphi+\omega\tau\}/2+\mathrm{h.c.}$
where $\mathcal{F}$ indicates a complex factor with the real and
imaginary parts
\begin{align}
\Re\{\mathcal{F}\} & =1-e^{-\Gamma(\tau)},\label{eq:Re_F}\\
\Im\{\mathcal{F}\} & =-e^{-\Gamma(\tau)/2}\sin\int_{\tau_{0}}^{\tau}\Omega(s)ds.\label{eq:Im_F}
\end{align}
In the two parts of the factor,
\begin{equation}
\Gamma(\tau)=\int_{\tau_{0}}^{\tau}ds\,\gamma(\Omega)\sin^{2}\varphi\label{eq:Gamma}
\end{equation}
converts the spectral function $\gamma(\Omega)$ into the time domain
to determine the effective decay in the response of the polarization.
It can be regarded as the bath-spectrum-weighted transform of Eq.~(\ref{eq:gamma})
and thus a decay factor corresponding to the bath correlations prescribed
in Eq.~(\ref{eq:correl}).

Equipped with the expression of $P(\tau)=P(t-x/v)$, we can determine
how the microwave pulse responds to the qubit and when its propagation
can be decoherence-free. Consider the standard Maxwell equation
\begin{equation}
\frac{\partial^{2}E}{\partial t^{2}}+\kappa c\frac{\partial E}{\partial t}-c^{2}\frac{\partial^{2}E}{\partial x^{2}}=-\frac{1}{\epsilon_{0}}\frac{\partial^{2}P}{\partial t^{2}}\label{eq:Maxwell_eqn}
\end{equation}
where $\kappa$ is the classical decay factor of the electric field.
Since $E(t)$ assumes the form of Eq.~(\ref{eq:elec_field}), in
which the envelop $\mathcal{E}(t)$ and the phase $\varphi(t)$ are
the slow variables compared to $\omega$, and the precession of $P(t)$
follows $\mathcal{F}(t)$, which is also slow compared to $\omega$,
the terms not on the order of $\omega$ can be ignored~\citep{mccall69,SCULLY}
after substituting the expressions of $E(t)$ and $P(t)$ into the
derivatives and Eq.~(\ref{eq:Maxwell_eqn}) can be linearized. Comparing
the coefficients of the carrier $e^{i(\varphi+\omega t-kx)}$ and
its conjugate, we obtain the coupled equations 
\begin{align}
\mathcal{E}\left(\frac{1}{v}-\frac{1}{c}\right)\frac{d\varphi}{d\tau} & =\frac{\mu k}{2\epsilon_{0}}\Re\{\mathcal{F}\},\label{eq:ph_eqn}\\
\left(\frac{1}{v}-\frac{1}{c}\right)\frac{d\mathcal{E}}{d\tau} & =-\frac{\mu k}{2\epsilon_{0}}\Im\{\mathcal{F}\},\label{eq:enve_eqn}
\end{align}
about the envelop and the phase, respectively, under the local time
frame $\tau=t-x/v$. In the equations, $v$ is the velocity of the
envelop wavefront and not necessarily equal to the phase velocity
$c$. The detailed derivations are given in Appendix B.

With Eqs.~(\ref{eq:ph_eqn}) and (\ref{eq:enve_eqn}), it becomes
clear that the factor $\mathcal{F}$ affects the envelop $\mathcal{E}(t)$
only through its imaginary part and the phase $\varphi(t)$ through
both its real and imaginary parts. For $\mathcal{E}(t)$, the effect
from \emph{$\mathcal{F}$} is the decoherence imposed by the environment
through the factor $e^{-\Gamma(t)/2}$. With the precedent sinusoidal
factor, decoherence would vanish when the integral $\int\Omega(\tau)d\tau$
is an integer multiple of $\pi$. Note that at qubit-pulse resonance,
$\Omega(t)$ reduces to $\mu\mathcal{E}(t)$. That means, omitting
the classical decay occurring in the waveguide owing to $\kappa$,
the envelop would retain its shape during the propagation as long
as the enveloping area of the pulse is an integer multiple of $\pi$.
This phenomenon exactly coincides with the observations of self-induced
transparency. In addition, \emph{$\mathcal{E}(t)$} not only preserves
its energy after traveling through the qubit, but makes the qubit
immune from the environment. That is, the decoherence factor vanishes
along with the sinusoidal factor, showing an $n\pi$-pulse would propagate
absorption-free and decoherence-free simultaneously. Consequently,
the master-Maxwell equation pair given by Eqs.~(\ref{eq:master_eqn})
and (\ref{eq:Maxwell_eqn}) serves as a microscopic foundation of
SIT effective for one single artificial atom and inclusive of environmental
effects.

\section{Decoherence-free propagation and ramification\label{sec:prop_and_ram}}

To find a general solution to Eq.~(\ref{eq:enve_eqn}) for $\mathcal{E}(t)$
with arbitrary initial area, we converts the integro-differential
equation of $\mathcal{E}$ into the second-order differential equation
\begin{equation}
\ddot{\mathcal{A}}=M^{2}e^{-\Gamma/2}\sin\mathcal{A}\label{eq:area_eqn}
\end{equation}
of the enveloped area $\mathcal{A}(\tau)=\mu\int_{\tau_{0}}^{\tau}ds\,\mathcal{E}(s)$
up to the wavefront, which is a pendulum equation augmented with a
decay factor. We have used $M=\sqrt{\mu^{2}kcv/2\epsilon_{0}(c-v)}$
to abbreviate the equation. Note that when the spectral function $\gamma(\Omega)$
happens to be orthogonal to $\sin^{2}\varphi$, $\Gamma$ vanishes
by definition, in which case even the dephasing reduces to zero according
to Eq.~(\ref{eq:Re_F}). However, for most models of thermal bath,
such as the ohmic and the sub-ohmic, the spectral function is an exponential
of $\Omega$ and thus would not make $\Gamma$ vanish in general.
If the qubit is initially assumed to be totally inverted with $\rho(0)=\left|e\right\rangle \left\langle e\right|$,
the diagonal elements of $\rho(t)$ determined by the master equation
(\ref{eq:master_eqn}) would only retain terms proportional to $e^{-\Gamma/2}$.
Therefore, we see that nonvanishing $\Gamma$ leads to finite dephasing
to the qubit as a dipole moment.

To find the analytic expression for $\mathcal{A}$, the pendulum equation
is first reduced to the first-order equation: $\dot{\mathcal{A}}=2Me^{-\Gamma/4}\sin(\mathcal{A}/2)$,
showing that $2n\pi$-pulses experience transparent transmission in
addition to being free from decoherence. For pulses of arbitrary enveloping
area, we retain the phase variable $\varphi$ in the expression of
$\Gamma(\tau)$ and solve Eq.~(\ref{eq:area_eqn}) formally as a
pendulum equation. The envelop $\mathcal{E}(\tau)$ as a time derivative
of $\mathcal{A}$ reads 
\begin{equation}
\mathcal{E}\left(\tau\right)=\frac{4M}{\mu}e^{-\Gamma(\tau)/4}\text{sech}M\left(\int_{\tau_{0}}^{\tau}ds\,e^{-\Gamma(s)/4}+\tau_{D}\right),\label{eq:area_sol}
\end{equation}
where $\tau_{D}$ is a delay time. Note that Eq.~(\ref{eq:area_sol})
retains the characteristic hypersecant hump of a soliton. The environment
culminates an attenuation on the pulse peak amplitude and a variable
time duration in the temporal argument, the latter of which affects
the traveling speed of pulses of different areas. The derivation process
of Eq.~(\ref{eq:area_sol}) is given in Appendix C.

\begin{figure}
\includegraphics[bb=15bp 0bp 385bp 260bp,clip,width=8.5cm]{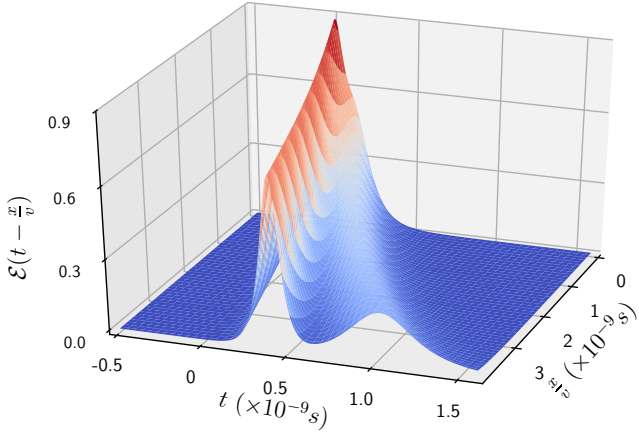}

\caption{Plot of the pulse envelop $\mathcal{E}(\tau)$ scale to arbitrary
unit as a function of local time $\tau=t-x/v$ , illustrating the
scenario of pulse ramification. The single pulse at the initial moment
$\tau=0$ splits into one $2\pi$-pulse traveling decoherence- and
absorption-free and one non-$n\pi$-pulse attenuating over time. System
parameters are taken from experiments of superconducting qubit circuits.~\label{fig:pulse_ram}}
\end{figure}

The manifestation of the environmental effects depends on the knowledge
of the spectral density $\gamma(\Omega)$ to determine the decay factor
$\Gamma(\tau)$. Without knowing its exact expression, we give a numerical
simulation of the $\mathcal{E}(\tau)$ evolution during the first
moments of qubit-pulse interaction by making the following assumptions:
(i) change of $\Omega(\tau)$ during the pulse-qubit interaction is
small relative to the bare qubit level spacing, thus adiabatic approximation
is valid; and (ii) the change of phase $\varphi(\tau)$ is small over
the course of interaction. The latter assumption is correlated to
the former, it will be evidence by the simulations to be given. It
follows that the integrand of $\Gamma(\tau)$ in Eq.~(\ref{eq:Gamma})
can be regarded as constant $4C_{0}$ under the slow variation, which
allows the approximation $\Gamma(\tau)\approx4C_{0}(\tau-\tau_{0})$
and leads to an exponential decay of the pulse peak in Eq.~(\ref{eq:area_sol})
(the scale factor 4 is added to simplify expressions below).

Under such premise, the integral in Eq.~(\ref{eq:area_sol}) is computed
numerically, giving rise to the propagation of a decaying pulse as
illustrated in Fig.~\ref{fig:pulse_ram} Note that $\mathcal{E}(\tau)$
under the local time frame $\tau$ would appear simply as a decaying
envelop peaking at $\tau=0$. To appreciate the propagation process
in the figure, we have returned the reference frame to the separate
laboratory axes $x/v$ and $t$, where parameters are set to values
accessible by typical qubits in superconducting circuits: $\omega=5$GHz
and a Q-factor of $10^{3}$~\citep{wen18,martinis05}. Since $\tau_{0}$
is an arbitrary initial time point, it is set to zero to simplify
the analysis.

We observe that a pulse of arbitrary enveloping area ramifies into
two: one of area a multiple of $2\pi$ travels freely, shown as one
ridge that converges to a constant height, and one of non-integral
area attenuates over $t$, shown as the other ridge in Fig.~\ref{fig:pulse_ram}.
Following the wavefronts of the two peaks, one observe that the slopes
of their projections onto the $x$-$t$ plane differs. The one traveling
decoherence-free pertains to a constant slope and therefore travels
at a constant velocity of light while the other has a curving slope.
The separation of wavefronts increases monotonically over time, showing
that the attenuating pulse is decelerating. This can be proved analytically
by taking the $\tau$-derivative of the argument of the hyper-secant
function in Eq.~(\ref{eq:area_sol}), giving the velocity $v\exp\{-C_{0}(t-t_{0})\}$. 

\begin{figure}
\includegraphics[bb=8bp 10bp 425bp 278bp,clip,width=8.5cm]{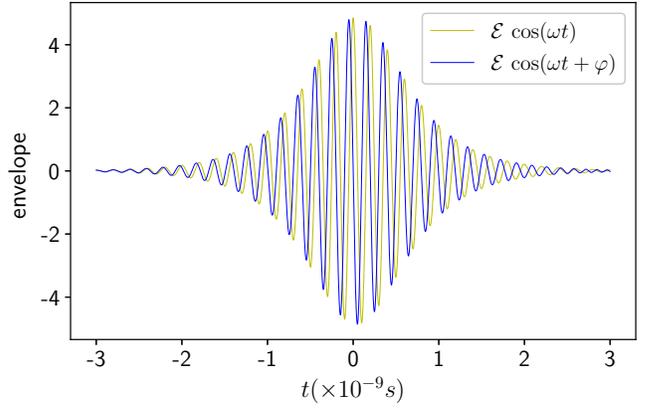}

\caption{The effect of the variation of pulse phase $\varphi(t)$ over time
is shown by the carrier wave under the typical hypersecant envelop
of a soliton pulse. The blue curve for a pulse with finite $\varphi(t)$
is set against a yellow curve for a constant phase and shows that
its oscillations are completed in advance to pulses free from environmental
effects.~\label{fig:phase}}
\end{figure}

The approximations taken in giving Fig.~\ref{fig:pulse_ram} is essentially
a first-order perturbative expansion of $\mathcal{E}(\tau)$, from
which the envelop and phase governed by Eqs.~(\ref{eq:ph_eqn})-(\ref{eq:enve_eqn})
are decoupled. Consequently, the dynamic phase accumulated when propagating
through the qubit is computed by integrating Eq.~(\ref{eq:ph_eqn}),
which reads
\begin{multline}
\varphi(\tau)=\varphi_{0}+M\int_{\tau_{0}}^{\tau}ds\,e^{-C_{0}(s-\tau_{0})}\sinh\left\{ 2C_{0}(s-\tau_{0})\right\} \\
\times\cosh\left\{ -\frac{M}{C_{0}}\left(e^{-C_{0}(s-\tau_{0})}-C_{0}\tau_{D}-1\right)\right\} \label{eq:phase}
\end{multline}
The second term contributed by the environment feedback generates
an advancement to the phase. Using the same system parameters as in
Fig.~\ref{fig:pulse_ram}, the integrals of Eq.~(\ref{eq:phase})
can be numerically computed, giving rise to the typical carrier wave
oscillations as plotted in Fig.~\ref{fig:phase}, where the phase
advancement over a duration of $20$ periods is shown against a carrier
of no phase variation.

\begin{figure}
\includegraphics[bb=10bp 10bp 425bp 280bp,clip,width=8.5cm]{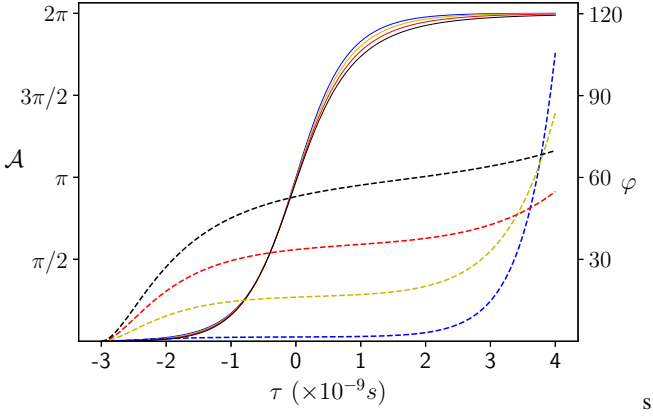}s

\caption{The pulse envelop area $\mathcal{A}(\tau)$ (solid curves and scaled
on the left axis) and the pulse carrier phase $\varphi(\tau)$ (dashed
curves scaled on the right axis) are plotted as functions of local
time $\tau$ over the same duration for the spectral densities $\gamma=4C_{0}=5$MHz
(blue curves), $50$MHz (yellow curves), $100$MHz (red curves), and
$150$MHz (black curves).~\label{fig:area_and_phase}}
\end{figure}

Since the three factors in the integrands of Eq.~(\ref{eq:phase})
are either exponential or variants of exponential functions, the phase
culminated on a pulse is a monotonically increasing function of time.
This is evidenced by the plots of $\varphi(\tau)$ given as dashed
curves in Fig.~\ref{fig:area_and_phase} for four different constant
spectral densities $\gamma$. The rate of phase accumulation increases
along with the increment of $\gamma$ when $\Gamma$, regarded as
a measure of rate of feedback from the environment is accordingly
increased. The envelop area $\mathcal{A}(\tau)$ computed from the
integral of $\mathcal{E}(\tau)$ in Eq.~(\ref{eq:area_sol}) is plotted
in the same figure, showing that the area variation accentuates on
the range of time where the phase variation is minimal, thereby ratifying
the assumptions we took above when arriving at the explicit solution
of $\mathcal{E}(\tau)$.

\section{Conclusion\label{sec:conclusion}}

To conclude, we have taken an adiabatic master equation approach to
analyze the propagation of a pulse through an environmentally coupled
qubit. The qubit would be free from decoherence when the pulse area
is $n\pi$. Further, when an othorgonal condition between the pulse
phase and the bath spectral density is satisfied, the dephasing vanishes
and thus relaxation times $T_{1}$ and $T_{2}$ become essentialy
infinite. It is also proved that when not vanishing the thermal environment
is responsible for inducing pulse ramifications which are frequently
observed in self-induced transparency experiments. The approach has
also enabled us for the first time to compute analytically the phase
variation in the pulse during its propagation through a two-level
system. Modeled on superconducting qubit circuits, the exact knowledge
on pulse-qubit interactions would benefit the designs of more sophisticated
microwave control pulses for quantum information processing.
\begin{acknowledgments}
Y.-B. Gao acknowledges the support of the National Natural Science
Foundation of China under Grant No.~11674017. H. I. acknowledges
the support by FDCT of Macau under grant 065/2016/A2, University of
Macau under grant MYRG2018-00088-IAPME, and National Natural Science
Foundation of China under grant No.~11404415.
\end{acknowledgments}

\appendix

\section{Adiabatic quantum master equation}

To construct a master equation for the system when taking into account
the geometric phases, we consider customarily a quantum heat bath
consisting of a multimode resonator $H_{\mathrm{B}}=\sum_{j}\omega_{j}a_{j}^{\dagger}a_{j}$
having dipole-field interaction $H_{I}=\sum_{j}g_{j}\left(a_{j}^{\dagger}+a_{j}\right)\sigma_{x}$
with the qubit.

Since the incident pulse with envelop $\mathcal{E}(x,t)$ is a weak
driving field to the qubit, the qubit evolution under the coupling
strength $\mu\mathcal{E}$ could be regarded as an adiabatic process
under Born-Oppenheimer approximation when compared to the thermal
decoherence process under couplings $\left\{ g_{j}\right\} $. Under
the total Hamiltonian
\begin{equation}
H(\tau)=H_{\mathrm{S}}^{\prime}(\tau)+H_{\mathrm{B}}+H_{\mathrm{I}}\label{total Hamiltonian}
\end{equation}
with $H_{\mathrm{S}}^{\prime}(\tau)+H_{\mathrm{B}}$ regarded as the
free energy, the adiabatic process is described by the master equation
for the universe $\mathscr{U}$: 
\begin{equation}
\frac{d\mathscr{U}}{d\tau}=-\int_{\tau_{0}}^{\tau}ds\left\{ \left[H_{I}(\tau),\left[H_{I}(\tau-s),\rho^{\prime}(\tau)\otimes\tilde{\rho}\right]\right]\right\} \label{eq:univ_eqn}
\end{equation}
in interaction picture. $\tilde{\rho}$ stands for the density matrix
for the thermal bath. The system-bath interaction $H_{I}(\tau)=U_{\mathrm{ad}}(\tau)H_{I}U_{\mathrm{ad}}^{\dagger}(\tau)$
has been transformed to the adiabatic evolution frame with the double-time
transformation $U_{\mathrm{ad}}(\tau-s)=e^{isH_{S}^{\prime}(\tau)}U_{\mathrm{ad}}(\tau)$
being used for the historic Hamiltonian $H_{I}(\tau-s)$ following
the convention~\citep{BREUER,albash12}.

As given in Eq.~(5) in the main text, the unitary transformations
involve the dressed basis defined at both the initial moment $\tau_{0}$
and the current momemt $\tau$ or the historic moment $\tau-s$. In
practice, we unify the basis reference to time $\tau$ by inversing
eigenstate definitions of Eqs.~(\ref{eq:nu_+})-(\ref{eq:nu_-}),
finding Eq.~(\ref{eq:U_ad}), i.e.
\begin{multline}
U_{\mathrm{ad}}(\tau)=\vert\nu_{+}(\tau)\rangle\langle\nu_{+}(\tau_{0})\vert e^{-i\phi_{+}(\tau)}\\
+\vert\nu_{-}(\tau)\rangle\langle\nu_{-}(\tau_{0})\vert e^{-i\phi_{-}(\tau)},
\end{multline}
based on which we can also derive
\begin{multline}
U_{\mathrm{ad}}(\tau-s)=e^{is\Omega_{+}(\tau)}\vert\nu_{+}(\tau)\rangle\langle\nu_{+}(\tau_{0})\vert e^{-i\phi_{+}}\\
+e^{is\Omega_{-}(\tau)}\vert\nu_{-}(\tau)\rangle\langle\nu_{-}(\tau)\vert e^{-i\phi_{-}}.\label{eq:hist_U_ad}
\end{multline}

During the adiabatic process, the bath variables $a_{j}$ and $a_{j}^{\dagger}$
stay relatively static while the operator $\sigma_{x}$ relevant to
the system is rotated. Therefore, the static $\sigma_{x}$ in the
dressed basis reads
\begin{align}
\sigma_{x}= & \left|e\right\rangle \left\langle g\right|+\left|g\right\rangle \left\langle e\right|\nonumber \\
= & -\cos\varphi(\tau)\vert\nu_{+}\rangle\langle\nu_{+}\vert+i\sin\varphi(\tau)\vert\nu_{+}\rangle\langle\nu_{-}\vert\nonumber \\
 & -i\sin\varphi(\tau)\vert\nu_{-}\rangle\langle\nu_{+}\vert+\cos\varphi(\tau)\vert\nu_{-}\rangle\langle\nu_{-}\vert.
\end{align}
Then following the qubit-field interaction using the unitary transformation
Eq.~(\ref{eq:hist_U_ad}), the operator during the interaction reads
for any historic moment $s$ at close resonance $(\Omega_{\pm}=\pm\Omega=\pm\mu\mathcal{E})$:
\begin{align}
\hat{\sigma}_{x}(\tau-s)= & U_{\mathrm{ad}}^{\dagger}(\tau-s)\sigma_{x}U_{\mathrm{ad}}(\tau-s)\nonumber \\
=-\cos\varphi(\tau) & U_{\mathrm{ad}}^{\dagger}(\tau)\left[\vert\nu_{+}\rangle\langle\nu_{+}\vert-\vert\nu_{-}\rangle\langle\nu_{-}\vert\right]U_{\mathrm{ad}}(\tau)\nonumber \\
+i\sin\varphi(\tau) & U_{\mathrm{ad}}^{\dagger}(\tau)\left[e^{-is\Omega}\vert\nu_{+}\rangle\langle\nu_{-}\vert-\mathrm{h.c.}\right]U_{\mathrm{ad}}(\tau),\label{eq:hist_sigma_x}
\end{align}
which includes the special case with $s=0$.

The derivation of the microscopic master equations begins with tracing
out the bath variables in the Liouville equation (\ref{eq:univ_eqn})
of the universe, giving
\begin{equation}
\frac{d\rho^{\prime}(\tau)}{dt}=-\int_{0}^{\tau-\tau_{0}}ds\mathrm{tr}_{\mathrm{B}}\left\{ \left[H_{\mathrm{I}}(\tau),\left[H_{\mathrm{I}}(\tau-s),\rho^{\prime}(\tau)\otimes\tilde{\rho}\right]\right]\right\} ,\label{eq:Liouville_2}
\end{equation}
where the integration limit is modified following Eq.~(\ref{eq:Liouville_eqn}).
We note that evoking the commutator generates the double-time operators
$\hat{\sigma}_{x}(\tau)\hat{\sigma}_{x}(\tau-s)$ according to Eq.~(\ref{eq:int_H}).
Therefore, equipped with Eq.~(\ref{eq:hist_sigma_x}), we compute
\begin{align}
 & \hat{\sigma}_{x}(\tau)\hat{\sigma}_{x}(\tau-s)=\nonumber \\
 & -\cos\varphi(\tau)U_{\mathrm{ad}}^{\dagger}(\tau)\sigma_{x}\left\{ \hat{\sigma}_{+}\hat{\sigma}_{-}-\hat{\sigma}_{-}\hat{\sigma}_{+}\right\} U_{\mathrm{ad}}(\tau)\nonumber \\
 & +\sin^{2}\varphi(\tau)U_{\mathrm{ad}}^{\dagger}(\tau)\left\{ e^{is\Omega}\hat{\sigma}_{+}\hat{\sigma}_{-}+e^{-is\Omega}\hat{\sigma}_{-}\hat{\sigma}_{+}\right\} U_{\mathrm{ad}}(\tau)\nonumber \\
 & -i\sin\varphi(\tau)\cos\varphi(\tau)U_{\mathrm{ad}}^{\dagger}(\tau)\left\{ e^{-is\Omega}\hat{\sigma}_{+}+e^{is\Omega}\hat{\sigma}_{-}\right\} U_{\mathrm{ad}}(\tau)\label{eq:db_time_cor}
\end{align}
and $\hat{\sigma}_{x}(\tau-s)\hat{\sigma}_{x}(\tau)=\left[\hat{\sigma}_{x}(\tau)\hat{\sigma}_{x}(\tau-s)\right]^{\dagger}$.
At the long-term limit $\tau\to\infty$, Eq.~(\ref{eq:Liouville_2})
expands into four terms. In dealing with the first term, we consider
\begin{align}
 & \int_{0}^{\infty}ds\mathrm{tr}_{\mathrm{B}}\left\{ H_{\mathrm{I}}(\tau)H_{\mathrm{I}}(\tau-s)\rho^{\prime}(\tau)\otimes\tilde{\rho}\right\} \nonumber \\
= & \intop ds\hat{\sigma}_{x}(\tau)\hat{\sigma}_{x}(\tau-s)\rho^{\prime}(\tau)\mathrm{tr}_{\mathrm{B}}\left\{ h_{\mathrm{B}}(\tau)h_{\mathrm{B}}(\tau-s)\tilde{\rho}\right\} \nonumber \\
= & \intop ds\hat{\sigma}_{x}(\tau)\hat{\sigma}_{x}(\tau-s)\rho^{\prime}(\tau)\sum_{j}g_{j}^{2}e^{-i\omega_{j}s}.
\end{align}
Note from the expression of Eq.~(\ref{eq:db_time_cor}) the double-time
operator product would contribute multiple terms in the integral,
but only those with the factor $e^{i(\Omega-\omega_{j})s}$ would
remain since the other terms containing the fast-oscillating exponential
factors would vanish after the integration over long period. Consequently,
since only this exponential factor involves in the integration over
the variable $s$, all other factors about time $\tau$ do not participate
in the integration and
\begin{equation}
\sum_{j}g_{j}^{2}\left[\intop dse^{i(\Omega-\omega_{j})s}\right]=\pi\sum_{j}g_{j}^{2}\delta(\Omega-\omega_{j})=\frac{1}{2}\gamma(\Omega)
\end{equation}
where the last equation employs the definition of Eq.~(\ref{eq:gamma}).
The integral would become
\begin{equation}
-\frac{1}{2}\gamma(\Omega)\sin^{2}\varphi U_{\mathrm{ad}}^{\dagger}\hat{\sigma}_{+}\hat{\sigma}_{-}U_{\mathrm{ad}}\rho^{\prime}
\end{equation}

When applying the same arguments to the other three terms in the expansion
of the RHS of Eq.~(\ref{eq:Liouville_2}), we arrives at 
\begin{multline}
\frac{d\rho^{\prime}}{d\tau}=-\gamma(\Omega)\sin^{2}\varphi\biggl[\frac{1}{2}\left\{ U_{\mathrm{ad}}^{\dagger}\hat{\sigma}_{+}\hat{\sigma}_{-}U_{\mathrm{ad}},\rho^{\prime}\right\} \\
-U_{\mathrm{ad}}^{\dagger}\hat{\sigma}_{-}U_{\mathrm{ad}}\rho^{\prime}U_{\mathrm{ad}}^{\dagger}\hat{\sigma}_{+}U_{\mathrm{ad}}\biggr].
\end{multline}
Finally, seeing that $\rho^{\prime}=U_{\mathrm{ad}}^{\dagger}\rho U_{\mathrm{ad}}$
is given in the interaction picture of the adiabatic evolution, we
note that for the conversion back to Schroedinger picture,
\begin{align}
\frac{d\rho}{d\tau} & =\frac{d}{d\tau}\left\{ U_{\mathrm{ad}}\rho^{\prime}U_{\mathrm{ad}}^{\dagger}\right\} \nonumber \\
 & =U_{\mathrm{ad}}\frac{d\rho^{\prime}}{d\tau}U_{\mathrm{ad}}^{\dagger}-iH_{\mathrm{S}}U_{\mathrm{ad}}\rho^{\prime}U_{\mathrm{ad}}^{\dagger}+iU_{\mathrm{ad}}\rho^{\prime}U_{\mathrm{ad}}^{\dagger}H_{\mathrm{S}}\nonumber \\
 & =-i\left[H_{\mathrm{S}},\rho\right]-\gamma(\Omega)\sin^{2}\varphi\left[\frac{1}{2}\left\{ \hat{\sigma}_{+}\hat{\sigma}_{-},\rho\right\} -\hat{\sigma}_{-}\rho\hat{\sigma}_{+}\right],
\end{align}
which is the Lindblad form of the master equation as in Eq.~(\ref{eq:master_eqn}).

\section{Equations of envelope and phase}

From the original Maxwell equation~(\ref{eq:Maxwell_eqn}), the loss
$\kappa$ in the waveguide is assumed negligible. Then the so-called
slowly-varying envelope approximation~\citep{basov66}, that is $\partial\mathcal{E}/\partial t\ll\omega\mathcal{E}$,
$\partial\mathcal{E}/\partial x\ll k\mathcal{E}$, $\partial\varphi/\partial t\ll\omega$,
and $\partial\varphi/\partial x\ll k$ for the electric field $E(t)$;
$\partial\mathcal{F}/\partial t\ll\omega\mathcal{F}$ and $\partial\mathcal{F}/\partial x\ll k\mathcal{F}$
for the polarization $P(t)$ can be taken. Thus when substituting
the expressions of $E(t)$ and $P(t)$ (i.e. in the form of Eq.~(\ref{eq:elec_field})),
the second-order partial derivates with respect to both time and space
are regarded as negligible terms in comparison to the linear terms
$\partial\mathcal{E}/\partial t$, etc. The Maxwell equation is essentially
reduced to a first-order PDE.

When comparing the real and the imaginary parts of both sides of this
PDE, one arrives at the coupled equations
\begin{align}
\frac{\partial\mathcal{E}}{\partial x}+\frac{1}{c}\frac{\partial\mathcal{E}}{\partial t} & =-\frac{k\mu}{2\epsilon_{0}}\Im\{\mathcal{F}\},\label{eq:PDE_E}\\
\frac{\partial\varphi}{\partial x}+\frac{1}{c}\frac{\partial\varphi}{\partial t} & =-\frac{k\mu}{2\epsilon_{0}}\frac{\Re\{\mathcal{F}\}}{\mathcal{E}},\label{eq:PDE_phi}
\end{align}
for the envelope variable and the phase variable. For the local time
$\tau=t-x/v$, the two derivative operators on the left hand side
can be combined, i.e.
\begin{align}
\frac{\partial}{\partial t}+c\frac{\partial}{\partial x} & =\frac{\partial}{\partial\tau}\frac{\partial\tau}{\partial t}+c\frac{\partial}{\partial\tau}\frac{\partial\tau}{\partial x}\nonumber \\
 & =\left(1-\frac{c}{v}\right)\frac{\partial}{\partial\tau},
\end{align}
under which Eqs.~(\ref{eq:PDE_E})-(\ref{eq:PDE_phi}) become ODEs
of one variable:
\begin{align}
\frac{d\mathcal{E}}{d\tau} & =\frac{\omega}{2\left(1-\frac{c}{v}\right)\epsilon_{0}}\Im\{\mathcal{F}\},\\
\frac{d\varphi}{d\tau} & =-\frac{\omega}{2\left(1-\frac{c}{v}\right)\epsilon_{0}}\frac{\Re\{\mathcal{F}\}}{\mathcal{E}},
\end{align}
where the equation of $\mathcal{E}(\tau)$ is decoupled from that
of $\varphi(\tau)$. Rearranging the factors on two sides lead to
Eqs.~(\ref{eq:ph_eqn})-(\ref{eq:enve_eqn}).

\section{Solving for envelope and phase}

Noticing that the sinusoidal factor in Eq.~(\ref{eq:Im_F}) is simply
$\sin\mathcal{A}$, given the definition of the envelope area $\mathcal{A}$,
we can write Eq.~(\ref{eq:area_eqn}) as
\begin{equation}
\frac{d\mathcal{E}}{d\tau}=-\frac{\mu ck}{2(c/v-1)\epsilon_{0}}e^{-\Gamma/2}\sin\mathcal{A}.
\end{equation}
Recognizing $d\mathcal{E}/d\tau=d^{2}\mathcal{A}/d\tau^{2}$ and using
the abbreviation $M=\sqrt{\mu^{2}kcv/2\epsilon_{0}(c-v)}$, we arrive
at the pendulum equation
\begin{equation}
\ddot{\mathcal{A}}=M^{2}e^{-\Gamma/2}\sin\mathcal{A}.
\end{equation}
Since $\ddot{\mathcal{A}}=d\dot{\mathcal{A}}/d\tau=\dot{\mathcal{A}}\left(d\dot{\mathcal{A}}/d\mathcal{A}\right)$,
the equation can be rewritten as
\begin{equation}
\dot{\mathcal{A}}d\dot{\mathcal{A}}=M^{2}e^{-\Gamma/2}\sin\mathcal{A}d\mathcal{A}.
\end{equation}
Formally integrating on both sides while regarding the decay factor
$\Gamma(\tau)$ as a slow-varying variable compared to \emph{$\mathcal{A}(\tau)$
}under a Born-Oppenheimer approximation, one gets
\begin{equation}
\dot{\mathcal{A}}^{2}=2M^{2}(1-\cos\mathcal{A})e^{-\Gamma/2}=4M^{2}e^{-\Gamma/2}\sin^{2}\frac{\mathcal{A}}{2}.
\end{equation}
Then taking the square root, one arrives at a first-order equation
\begin{equation}
\frac{d\mathcal{A}}{d\tau}=2Me^{-\Gamma/4}\sin\frac{\mathcal{A}}{2},\label{eq:1or_area_eqn}
\end{equation}
whereby the $\sin\mathcal{A}/2$ factor can be moved to LHS and we
can formally solve for $\mathcal{A}$:
\begin{equation}
\ln\tan\frac{\mathcal{A}}{4}=M\int_{\tau_{0}}^{\tau}e^{-\Gamma/4}ds+C
\end{equation}
where $\tau_{0}$ is an arbitrary initial time of integration and
$C$ is the integration constant. Hence,
\begin{equation}
\tan\frac{\mathcal{A}}{4}=C\exp\left\{ M\int_{\tau_{0}}^{\tau}e^{-\Gamma/4}ds\right\} 
\end{equation}
and letting $\mathcal{A}_{0}=\mathcal{A}(\tau_{0})$ shows $C=\tan\mathcal{A}_{0}/4$.
Abosrbing $C$ into the exponential, we can simplify the above into
\begin{equation}
\tan\frac{\mathcal{A}}{4}=\exp\left\{ M\int_{\tau_{0}}^{\tau}e^{-\Gamma/4}ds+\tau_{D}\right\} 
\end{equation}
where $\tau_{D}=\frac{1}{M}\ln\tan\mathcal{A}_{0}/4$ can be regarded
as the delay time. Then, using the identity $\sin2\theta=2\left[\tan\theta+\left(\tan\theta\right)^{-1}\right]^{-1}$,
Eq.~(\ref{eq:1or_area_eqn}) can be written as
\begin{equation}
\frac{d\mathcal{A}}{d\tau}=4Me^{-\Gamma/4}\mathrm{sech}\left\{ M\int_{\tau_{0}}^{\tau}e^{-\Gamma/4}ds+\tau_{D}\right\} ,
\end{equation}
which gives Eq.~(\ref{eq:area_sol}).

For the phase $\varphi(\tau)$, we substitute the decay factor Eq.~(\ref{eq:Re_F})
into Eq.~(\ref{eq:ph_eqn}) to get
\begin{equation}
\frac{d\varphi}{d\tau}=\frac{\mu ck}{2\mathcal{E}(c/v-1)\epsilon_{0}}\left(1-e^{-\Gamma}\right)=\frac{M^{2}}{\mu\mathcal{E}}\left(1-e^{-\Gamma}\right),\label{eq:intm_ph}
\end{equation}
which is an integro-differential equation, considering the expression
of $\mathcal{E}$ in Eq.~(\ref{eq:area_sol}). Like described in
the text, we simplify the consideration by reducing $\Gamma(\tau)$
to a linear dependence on time, writting $\Gamma(\tau)=4C_{0}(\tau-\tau_{0})$
and hence
\begin{multline}
\int_{\tau_{0}}^{\tau}e^{-\Gamma(s)/4}ds=\int_{\tau_{0}}^{\tau}e^{-C_{0}(s-\tau_{0})}ds\\
=-\frac{1}{C_{0}}\left[e^{-C_{0}(\tau-\tau_{0})}-1\right].
\end{multline}
Therefore, Eq.~(\ref{eq:intm_ph}) reads
\begin{align}
\frac{d\varphi}{d\tau}= & \frac{M}{4}\frac{e^{\Gamma/4}-e^{-3\Gamma/4}}{\text{sech}M\left(\int_{\tau_{0}}^{\tau}ds\,e^{-\Gamma(s)/4}+\tau_{D}\right)}\nonumber \\
= & \frac{M}{4}\left[e^{C_{0}(\tau-\tau_{0})}-e^{-3C_{0}(\tau-\tau_{0})}\right]\times\nonumber \\
 & \cosh\left\{ -\frac{M}{C_{0}}\left(e^{-C_{0}(\tau-\tau_{0})}-C_{0}\tau_{D}-1\right)\right\} .
\end{align}
Then integrating both sides with respect to the local time $\tau$
yields Eq.~(\ref{eq:phase}).


\begin{thebibliography}{99}
\bibitem{wallraff04}A. Wallraff, D. I. Schuster, A. Blais, L. Frunzio,
R.-S. Huang, J. Majer, S. Kumar, S. M. Girvin, and R. J. Schoelkopf,
Nature \textbf{431}, 162 (2004).

\bibitem{blais04}A. Blais, R.-S. Huang, A. Wallraff, S. M. Girvin,
and R. J. Schoelkopf, Phys. Rev. A \textbf{69}, 062320 (2004).

\bibitem{neeley10}M. Neeley, R. C. Bialczak, M. Lenander, E. Lucero,
M. Mariantoni, A. D. O\textquoteright Connell, D. Sank, H. Wang, M.
Weides, J. Wenner, Y. Yin, T. Yamamoto, A. N. Cleland, and J. M. Martinis,
Nature \textbf{467}, 570 (2010).

\bibitem{eichler12}C. Eichler, C. Lang, J. M. Fink, J. Govenius,
S. Filipp, and A. Wallraff, Phys. Rev. Lett. \textbf{109}, 240501
(2012).

\bibitem{marian11}M. Mariantoni, H. Wang, T. Yamamoto, M. Neeley,
R. C. Bialczak, Y. Chen, M. Lenander, E. Lucero, A. D. O\textquoteright Connell,
D. Sank, M. Weides, J. Wenner, Y. Yin, J. Zhao, A. N. Korotkov, A.
N. Cleland, and J. M. Martinis, Science \textbf{334}, 61 (2011).

\bibitem{lucero12}E. Lucero, R. Barends, Y. Chen, J. Kelly, M. Mariantoni,
A. Megrant, P. O\textquoteright Malley, D. Sank, A. Vainsencher, J.
Wenner, T. White, Y. Yin, A. N. Cleland, and J. M. Martinis, Nat.
Phys. \textbf{8}, 719 (2012).

\bibitem{mallet09}F. Mallet, F. R. Ong, A. Palacios-Laloy, F. Nguyen,
P. Bertet, D. Vion, and D. Esteve, Nat. Phys. \textbf{5}, 791 (2009).

\bibitem{dicarlo10}L. DiCarlo, M. D. Reed, L. Sun, B. R. Johnson,
J. M. Chow, J. M. Gambetta, L. Frunzio, S. M. Girvin, M. H. Devoret,
and R. J. Schoelkopf, Nature \textbf{467}, 574 (2010).

\bibitem{motzoi09}F. Motzoi, J. M. Gambetta, P. Rebentrost, and F.
K. Wilhelm, Phys. Rev. Lett. \textbf{103}, 110501 (2009).

\bibitem{chow10}J. M. Chow, L. DiCarlo, J. M. Gambetta, F. Motzoi,
L. Frunzio, S. M. Girvin, and R. J. Schoelkopf, Phys. Rev. A \textbf{82},
040305 (2010).

\bibitem{lidar98}D. A. Lidar, I. L. Chuang, and K. B. Whaley, Phys.
Rev. Lett. \textbf{81}, 2594 (1998).

\bibitem{gambetta11}J. M. Gambetta, A. A. Houck, and A. Blais, Phys.
Rev. Lett. \textbf{106}, 030502 (2011).

\bibitem{koch07}J. Koch, T. M. Yu, J. Gambetta, A. A. Houck, D. I.
Schuster, J. Majer, A. Blais, M. H. Devoret, S. M. Girvin, and R.
J. Schoelkopf, Phys. Rev. A \textbf{76}, 042319 (2007).

\bibitem{schreier08}J. A. Schreier, A. A. Houck, J. Koch, D. I. Schuster,
B. R. Johnson, J. M. Chow, J. M. Gambetta, J. Majer, L. Frunzio, M.
H. Devoret, S. M. Girvin, and R. J. Schoelkopf, Phys. Rev. B \textbf{77},
180502 (2008).

\bibitem{ian10}H. Ian, Y.-X. Liu, and F. Nori, Phys. Rev. A \textbf{81},
063823 (2010).

\bibitem{wen18}P. Y. Wen, A. F. Kockum, H. Ian, J. C. Chen, F. Nori,
and I.-C. Hoi, Phys. Rev. Lett. \textbf{120}, 063603 (2018).

\bibitem{jqyou11}J. Q. You and F. Nori, Nature \textbf{474}, 589
(2011).

\bibitem{lamb71} G. L. Lamb, Rev. Mod. Phys. \textbf{43}, 99 (1971).

\bibitem{lamb73}G. L. Lamb, Phys. Rev. Lett. \textbf{31}, 196 (1973).

\bibitem{mccall67} S. L. McCall and E. L. Hahn, Phys. Rev. Lett.
\textbf{18}, 908 (1967).

\bibitem{mccall69}S. L. McCall and E. L. Hahn, Phys. Rev. \textbf{183},
457 (1969). 

\bibitem{slusher72} R. E. Slusher and H. M. Gibbs, Phys. Rev. A \textbf{5},
1634 (1972).

\bibitem{basov66}N. G. Basov, R. V. Ambartsumyan, V. S. Zuev, P.
G. Kryukov, and V. S. Letokhov, Soviet JETP \textbf{23}, 16 (1966).

\bibitem{wilson07}C. M. Wilson, T. Duty, F. Persson, M. Sandberg,
G. Johansson, and P. Delsing, Phys. Rev. Lett. \textbf{98}, 257003
(2007). 

\bibitem{BREUER}H.-P. Breuer and F. Petruccione, \emph{The Theory
of Open Quantum Systems} (Oxford University, 2002).

\bibitem{albash12}T. Albash, S. Boixo, D. A. Lidar, and P. Zanardi,
New J. Phys. \textbf{14}, 123016 (2012).

\bibitem{SCULLY}M. O. Scully and M. S. Zubairy, \emph{Quantum Optics}
(Cambridge University Press, 1997).

\bibitem{martinis05}J. M. Martinis, K. B. Cooper, R. McDermott, M.
Steffen, M. Ansmann, K. D. Osborn, K. Cicak, S. Oh, D. P. Pappas,
R. W. Simmonds, and C. C. Yu, Phys. Rev. Lett. \textbf{95}, 210503
(2005). 
\end{thebibliography}
\end{document}